\documentclass[sigconf, authorversion=true]{acmart}
\usepackage{microtype}

\AtBeginDocument{%
  \providecommand\BibTeX{{%
    \normalfont B\kern-0.5em{\scshape i\kern-0.25em b}\kern-0.8em\TeX}}}

\setcopyright{acmlicensed}
\acmPrice{15.00}
\acmDOI{10.1145/3397537.3397552}
\acmYear{2020}
\copyrightyear{2020}
\acmSubmissionID{prog20eniac-id3-p}
\acmISBN{978-1-4503-7507-8/20/03}
\acmConference[<Programming'20> Companion]{Companion Proceedings of the 4th International Conference on the Art, Science, and Engineering of Programming}{March 23--26, 2020}{Porto, Portugal}
\acmBooktitle{Companion Proceedings of the 4th International Conference on the Art, Science, and Engineering of Programming (<Programming'20> Companion), March 23--26, 2020, Porto, Portugal}

\begin{document}

\title[Achieving Guidance in Applied Machine Learning]{Achieving Guidance in Applied Machine Learning \\through Software Engineering Techniques}

\author{Lars Reimann}
\email{reimann@cs.uni-bonn.de}
\author{G\"unter Kniesel-W\"unsche}
\email{gk@cs.uni-bonn.de}
\affiliation{%
  \institution{Smart Data Analytics, University of Bonn}
  \streetaddress{Endenicher Allee 19C}
  \city{Bonn}
  \country{Germany}
  \postcode{53859}
}


\begin{abstract}
  Development of machine learning (ML) applications is hard. Producing successful applications requires, among others, being deeply familiar with a variety of complex and quickly evolving application programming interfaces (APIs). It is therefore critical to understand what prevents developers from learning these APIs, using them properly at development time, and understanding what went wrong when it comes to debugging. We look at the (lack of) \emph{guidance} that currently used development environments and ML APIs provide to developers of ML applications, contrast these with software engineering best practices, and identify gaps in the current state of the art. We show that current ML tools fall short of fulfilling some basic software engineering gold standards and point out ways in which software engineering concepts, tools and techniques need to be extended and adapted to match the special needs of ML application development. Our findings point out ample opportunities for research on ML-specific software engineering.
\end{abstract}

\ccsdesc[300]{Software and its engineering~Software verification and validation}
\ccsdesc[300]{Software and its engineering~Automated static analysis}
\ccsdesc[500]{Computing methodologies~Machine learning}
\ccsdesc[300]{Software and its engineering~Software libraries and repositories}
\ccsdesc[300]{Software and its engineering~General programming languages}
\ccsdesc[500]{Software and its engineering}

\keywords{machine learning, software engineering, guidance, usability, learnability}

\maketitle

\section{Introduction} 
\label{sec:introduction}

The success of deep learning and machine learning (ML) in general in recent years is attributed to the availability of large datasets, more efficient algorithms, and specialized hardware \cite{Chollet2017}. Based on the current hype, more and more companies try to integrate ML into their products, leading to an ever increasing demand of data scientists. Software developers who already know the targeted application domain could fill this gap but typically lack a scientific ML background. Therefore, they need \emph{guidance} that helps them correctly use ML.

We define guidance to encompass all means to facilitate or enforce correct usage of a tool or an API or proper adherence to a workflow. Guidance includes, on one hand, preventing, detecting, explaining and fixing erroneous usage and, on the other hand, communicating best practices and helping apply them. In a traditional software engineering (SE) context, guidance can help developers when learning an API or the use of a tool, developing a program using the API or tool, and debugging the program when something went wrong.

This paper analyses how current ML tools and APIs fail to provide guidance to developers and presents ideas for improvement. 
Sec. \ref{sec:workflows} reviews a typical workflow of ML application development, points out differences to traditional software engineering, and elaborates the desirable guidance for each step of the ML workflow. 
Sec. \ref{sec:stateOfArt} reviews how the main APIs, languages, and development environments currently used for ML application development fail to provide guidance. 
Sec. \ref{sec:simpleml} discusses how better guidance can be offered by applying and adapting various SE concepts to the specific requirements of ML application development.

\section{ML Workflows are Different}
\label{sec:workflows}

Due to space constraints, we focus without loss of generality on supervised learning, the currently most widely-used ML workflow \cite{Chollet2017}. The task in supervised learning is to learn a function $H$ on the basis of some \emph{training data}, which is a (typically very large) set of examples containing \emph{features} and \emph{labels}. Given the features, $H$ should ideally produce the labels. If the labels are from a continuous set, we call it a \emph{regression} problem, if the labels are from a discrete set we call it a \emph{classification} problem.

A traditional SE workflow involves activities such as requirement engineering, design, implementation, refactoring, and testing. In an ML workflow, most of these activities are performed in a quite specific way (e.g. Sec. \ref{subsec:requirements} and \ref{subsec:tdd}), if at all\footnote{What, for instance, is the equivalent of refactoring in ML?}. Moreover, an ML workflow involves additional essential activities, such as Data Engineering (Sec. \ref{subsec:data_engineering}), Model Engineering (Sec. \ref{subsec:model_engineering}), and Model Quality Engineering (Sec. \ref{subsec:quality_engineering}).

\subsection{Evaluation-Focussed Requirements Engineering}
\label{subsec:requirements}
Requirements Engineering (RE) for ML reflects the fact that ML development includes empirical evaluation of results as part of the indispensable Model Quality Engineering activity (Sec. \ref{subsec:quality_engineering}). Accordingly, the definition of quality metrics and the required level of quality for each metric is not an optional best practice but an essential part of the workflow. One quality metric we might define upfront, for instance, is that at least $80\%$ accuracy must be achieved. This gives as clear metric to assess the usefulness of a model and a stopping criteria for the workflow.

\paragraph{Requirements Engineering Guidance}

Desirable guidance for RE for ML includes:
\begin{itemize}
    \item Automated suggestions of suitable quality metrics based on problem (regression / classification) and data.
    \item Automated checking if a comparable task has been solved before and achieved the desired quality.
\end{itemize}

\subsection{Test-Driven Development the ML Way}
\label{subsec:tdd}

Before we can start learning, we must first ensure that a large amount of data is available and split it into a  
\begin{itemize}
        \item a \emph{training set} that will be used for the actual learning / training (Sec. \ref{subsec:model_engineering}),
        \item a \emph{validation set} that will be used for assessing whether the training results are good enough and to fine-tune the model\footnote{Alternatively, we can use \emph{cross-validation} \cite{Chollet2017}.} (Sec. \ref{subsec:quality_engineering}).
        \item a \emph{test set}, that will be used to get a last measure of how well the model performs on real-world data before pushing the model into production (Sec. \ref{subsec:quality_engineering}). 
\end{itemize}

\paragraph{Test-Driven Development Guidance}
Desirable guidance for Test-Driven Development in ML includes:
\begin{itemize}
    \item Automated suggestion whether to use cross-validation or a specific validation set based on desired quality, training time and the amount of data available.
    \item Automated suggestion of what percentage of the data should go into each of the sets.
    \item Automated verification of best practices for a `good' split of the data. For instance, the distribution of labels and significant features in the three sets should be as similar as possible (\emph{stratified sampling} \cite{Geron2017}).
    \item Automatically hiding the test set away until needed, so it does not influence decisions in subsequent steps of the workflow like the choice of the model \cite{Chollet2017}.
\end{itemize}

\subsection{Data Engineering}
\label{subsec:data_engineering}
Next, we must ensure that the available training data is of sufficient quality. If not, we need to perform \emph{data engineering}, which involves 
     \begin{itemize}
        \item handling of missing values,
        \item turning categorical into numerical values,
        \item normalization of numerical values, 
        \item assessing the statistical relevance of available features,
        \item selecting the features used for learning, possibly combining several features.
 	\end{itemize}
Data engineering is typically the most time consuming step in an ML project but also the most important one: If the data used for learning is of poor quality, so will be the learned function $H$. 

\paragraph{Data Engineering Guidance} 
Desirable guidance for data engineering includes:
\begin{itemize}
        \item Automated detection of missing values and automated suggestions for handling them, ideally specific to the type of data or learning task at hand. For instance, for numeric values, missing values could be substituted by the mean of the other values. This is not possible for a feature that represents street names.   
        \item Automatic conversion of categorical into numerical values, or at least suggestions for possible conversions, again taking into account the specific data types and application semantics. For instance, if categorical values can be ordered (e.g. `poor', `middle class', `rich', `billionaire'), we can replace them by an increasing sequence of numbers (\emph{label encoding}). Otherwise, we can perform a \emph{one-hot encoding}, which maps each different value to a sparse vector \cite{Geron2017}.
        \item Automated normalization of numerical values. 
        \item Automated computation of the entropy of the distribution of a feature or of any other measure for the statistical relevance of the available features.
        \item Automated suggestions for sensible feature combinations in a specific application domain.
        \item Ideally, any data processing step performed on the training data, should be automatically applied also to the validation and test set in order to avoid inconsistencies or tedious and error-prone manual application of each step on each dataset.  
 \end{itemize}

\subsection{Model Engineering}
\label{subsec:model_engineering}
The next step in the ML workflow is the selection of an existing ML API and the choice of a supported \emph{learning algorithm}, such as \emph{decision trees} or \emph{neural networks} (NNs). At first glance, this might appear similar to the choice of proper algorithms and data structures in SE. However, it was shown by Wolpert in \cite{Wolpert1996} that lacking specific assumptions no learning algorithm performs better than any other. This makes it impossible to provide a general guideline to always use, say NNs.

Choosing a learning algorithm instead requires significant background in ML and a careful consideration of the application goal\footnote{Do we need explainability?} and the properties of the data available after data engineering. Even with this knowledge, though, the final choice of the proper learning algorithm is typically the outcome of a lot of experimentation, involving several iterations of the entire ML development workflow.

Given a choice of an algorithm, a developer then writes a \emph{training program} and starts the learning process on the prepared training data. Depending on the circumstances (hardware, data size, learning algorithm, etc.) this step can take a few seconds or several days. Its outcome is a \emph{model} representing the learned function $H$. It is worth noting that, unlike in SE, the learned model is the final output of the workflow, not the written program.

\paragraph{Model Engineering Guidance}
Desirable guidance for model engineering includes:
\begin{itemize}
        \item Automated suggestions of suitable learning algorithms based on the available data and the application domain.
        \item Well-documented APIs and all kinds of teaching and training material that help quickly learn how to use them.  
        \item Automated checking of correct API use, ideally during program development (static checking) but at least at runtime. Given that a single training run can take hours and days and the entire workflow involves many iterations, the importance of static checking cannot be overstated. Costs incurred in statically detecting and fixing errors in the training program that would be prohibitive in traditional SE appear small in an ML context. 
        \item Help in understanding and fixing errors, ranging from sensible error messages, ideally enriched with suggestions how to resolve the error, to automated quickfixes.
 \end{itemize}

\subsection{Model Quality Engineering}
\label{subsec:quality_engineering}   
Next, the performance of the model is evaluated by running it on the validation set (Sec. \ref{subsec:tdd}), to get a measure of how well the model performs on data not seen during training. By measuring the metrics established in the RE phase (Sec. \ref{subsec:requirements}) we check that the model learned the characteristics of the training data well enough (no \emph{underfitting}) and did not simply learn the training data by heart (no \emph{overfitting}). The gathered metrics are used to fine-tune the training program by setting different \emph{hyperparameters} that control the behaviour of the learning algorithm. For example, the maximum number of nodes in a decision tree can be a hyperparameter. A low value can lead to underfitting, while a high value can lead to overfitting. With the improved settings, a new training is started and the entire process is repeated from there. If the results are still not satisfactory, one might go further back to select a different learning algorithm (Sec. \ref{subsec:model_engineering}) or even rethink and enhance data engineering (Sec. \ref{subsec:data_engineering}).

Finally, when the model has passed the validation stage, we use the test set (Sec. \ref{subsec:tdd})  to get a last measure of how well the model performs on unseen data before pushing the model into production. We cannot use the validation set for this, since we selected the model that works best on the validation set, which introduces a risk of overfitting \cite{Geron2017}. 

\paragraph{Model Quality Engineering Guidance}
Desirable guidance for quality engineering includes:
\begin{itemize}
        \item Automated experiment management, that is, saving of the $n$ models that achieved the highest performance values during validation along with the related training program and fully data-engineered datasets (including training, validation and test data) in order to support reproducibility of results.    
        \item Automated detection of overfitting and related suggestions how to combat it, e.g. by inserting drop-out layers \cite{Hinton2012} in a neural network.  
        \item Automated detection of underfitting and related suggestions how to combat it, e.g. by choosing a more powerful learning algorithm.  
 \end{itemize}

\section{State of the Art}
\label{sec:stateOfArt}

The highly explorative and iterative workflow described in Sec. \ref{sec:workflows} is supported by high-level APIs for ML (Sec. \ref{subsec:api}), dynamic languages (Sec. \ref{subsec:languages}), and dynamic development environments (Sec. \ref{subsec:ide}). In this section, we review each of them with respect to the guidance they provide.


\subsection{Efficient, High-Level APIs} 
\label{subsec:api}
Based on a survey of ML APIs by Nguyen and others from 2019 \cite{Nguyen2019} we investigated scikit-learn \cite{Pedregosa2011}, a representative of an API for `classical' ML, as well as Keras \cite{chollet2015keras} and PyTorch \cite{Paszke2019}, which are APIs for deep learning. These APIs provide high-level abstractions to apply and adapt learning algorithms (model engineering), to ensure sufficient prediction quality (model quality engineering) and even to perform data engineering, thereby providing all functionality needed to rapidly iterate the ML workflow from Sec. \ref{sec:workflows}. With respect to guidance, though, we identified several shortcomings:

\paragraph{Criticism: Hidden and Inconsistent Constraints in the API Documentation} The broadness of these APIs makes it difficult to use them correctly, especially since many constraints are hard to understand from the documentation. By looking just at the documentation of the support vector machine learning algorithm for classification (SVC) in scikit-learn\footnote{\url{https://scikit-learn.org/stable/modules/generated/sklearn.svm.SVC.html}} we can already identify four different categories of constraints documented only in natural language or documented inconsistently:
\begin{itemize}
	\item \emph{type constraints} limit values of expressions in any context,
	\item \emph{dependencies} limit values depending on other values,
	\item \emph{temporal constraints} limit the order of operations, 
	\item \emph{execution context constraints} express restrictions specific to the used programming language or executions platform. 
\end{itemize}

\begin{figure}[H]
    \centering
    \captionsetup{justification=centering}
    \fbox{\includegraphics[width=0.8\linewidth]{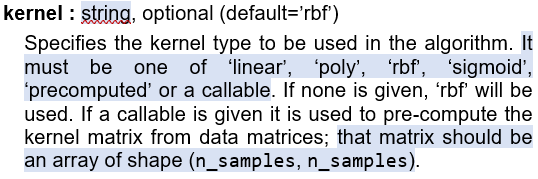}}
    \caption{Type constraints (scikit-learn)}
    \label{fig:scikit_docs_kernel}
\end{figure}

Hidden and inconsistent \emph{type constraints} are illustrated in Fig. \ref{fig:scikit_docs_kernel}: According to the first line of the  documentation, the constructor parameter \verb|kernel| of the SVC must be of type string. However, the rest of the text (see highlighting), reveals that only the five string literals `linear', `poly', `rbf', `sigmoid' and `precomputed' are allowed and that we can additionally pass a \emph{callable}, thus something that is not a string. Reading on, we learn that the signature of the callable is further restricted: It must have a single parameter that must be an array of shape n\_samples by n\_samples. 

\begin{figure}[H]
    \centering
    \captionsetup{justification=centering}
    \fbox{\includegraphics[width=0.8\linewidth]{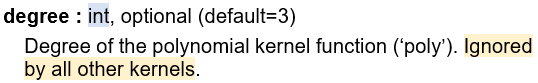}}
    \caption{Dependent API elements (scikit-learn)}
    \label{fig:scikit_docs_degree}
\end{figure}

A \emph{dependency constraint} is shown in Fig. \ref{fig:scikit_docs_degree}: The value of the constructor parameter \verb|degree| is only relevant if the constructor parameter \verb|kernel| has the value `poly'.

\begin{figure}[H]
    \centering
    \captionsetup{justification=centering}
    \fbox{\includegraphics[width=0.8\linewidth]{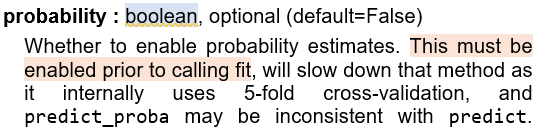}}
    \caption{Temporal constraint (scikit-learn)}
    \label{fig:scikit_docs_probabilty}
\end{figure}

A \emph{temporal constraint} is shown in Fig. \ref{fig:scikit_docs_probabilty}: We must set the attribute \verb|probability| to `True' prior to the first call of the \verb|fit| method (which starts the learning process).

\begin{figure}[H]
    \centering
    \captionsetup{justification=centering}
    \fbox{\includegraphics[width=0.8\linewidth]{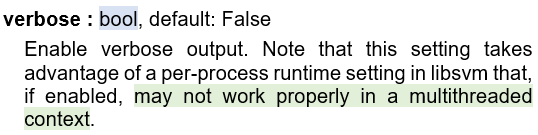}}
    \caption{Constraint of execution context (scikit-learn)}
    \label{fig:scikit_docs_verbose}
\end{figure}

An \emph{execution context constraint} is shown in Fig. \ref{fig:scikit_docs_verbose}: Verbose output may not work in a multithreaded environment.

\paragraph{Criticism: Lack of Constraint Checking and Helpful Error Messages}
Let us now investigate what happens if we run a program that does not comply with some of the restrictions shown above. We begin with the simple program in List. \ref{fig:scikit_errors_no_error}. There, we create an SVC with the invalid kernel `line' (see Fig. \ref{fig:scikit_docs_kernel}) --- the programmer probably meant 'linear'. But despite the wrong value, the code executes without error. The programmer therefore adds more code to preprocess the data and initiate training, which leads to the program state in List. \ref{fig:scikit_errors_bad_error_program}. Executing this program throws the error shown in List. \ref{fig:scikit_errors_bad_error_output}: 

\setcounter{figure}{0}
\renewcommand{\figurename}{Listing}

\begin{figure}[H]
    \centering
    \captionsetup{justification=centering}
    \fbox{\includegraphics[width=0.8\linewidth]{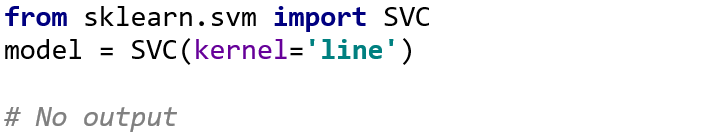}}
    \caption{Incorrect, but throws no error (scikit-learn)}
    \label{fig:scikit_errors_no_error}
\end{figure}

\begin{figure}[H]
    \centering
    \captionsetup{justification=centering}
    \fbox{\includegraphics[width=0.8\linewidth]{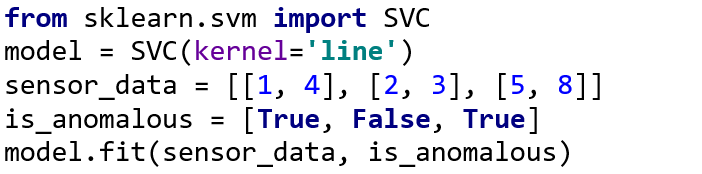}}
    \caption{Throws a misleading error (scikit-learn)}
    \label{fig:scikit_errors_bad_error_program}
\end{figure}

\begin{figure}[H]
    \centering
    \captionsetup{justification=centering}
    \fbox{\includegraphics[width=0.85\linewidth]{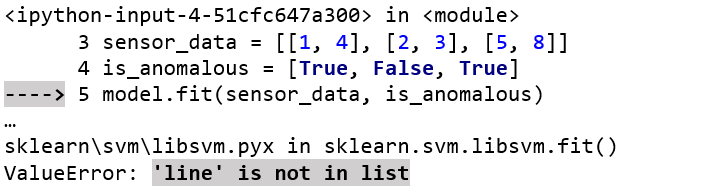}}
    \caption{Error thrown by program in List. \ref{fig:scikit_errors_bad_error_program} (scikit-learn)}
    \label{fig:scikit_errors_bad_error_output}
\end{figure}

\setcounter{figure}{4}
\renewcommand{\figurename}{Figure}

There are multiple issues with this behaviour: 
\begin{itemize}
	\item \emph{No precondition checking in API function.} First, the error is not thrown by the API at the place where the illegal value was set, misleading the programmer to think that the first two lines are correct. 
	\item \emph{Misleading, generic error messages.} A generic error is thrown by the runtime system when stumbling over the illegal value, which is much too late. At this point, the error message and shown stacktrace further mislead the programmer. Since lines 3 and 4 contain Python lists that indeed do not contain the string `line', the programmer could try to find a bug there. Going back to line 2, which apparently executed flawlessly, requires deep internal knowledge of the API, which is unlikely for anybody except its authors. One could argue that the string literal `line' shows up in line 2, which could point the programmer to the true issue. But note that the code for data preprocessing is normally much longer than shown here, so the line where the error originated might not be visible in the context where it manifests itself.
	\item \emph{No hint how to recover.} Finally, once the programmer has found the true bug, he must consult the documentation to find the correct value since the generic error message does not specify the valid inputs. 
\end{itemize}

\paragraph{Criticism: No Checking of ML Best Practices} We only looked at constraints of the API so far. For general ML best practices (Sec. \ref{sec:workflows}), the situation is even more dire: For them none of the APIs provides any checking or feedback. 
In addition, due to the differences between the APIs, each one would have to implement checking of these best practices separately, even though they apply regardless of the API that is used.

\paragraph{Criticism: No Static Checking} In either case, even the best documentation and ideal runtime error checking still requires developers to wait for (possibly long) training runs just to find out that an error occured. The effect is an annoying loop of editing, running, debugging, fixing and re-running the training program in order to see if the fix was successful --- the runtime acts like a test oracle. Ideally, the developer would get guidance while he writes the program but this is poorly supported in the programming languages used for ML development, which brings us to the next point.

\subsection{Languages}
\label{subsec:languages}
The above-mentioned study by Nguyen, et al. from 2019 \cite{Nguyen2019} found that Python is the most popular language for ML. This can be attributed to its readability and to the fact that it is an interpreted language and, thus, well suited for highly interactive and explorative development. Moreover, Python has a vast ecosystem of scientific APIs that are useful for ML. Being a general-purpose programming language, Python provides the flexibility needed to implement new learning algorithms and highly customized ML-workflows. However, much of this flexibility --- and therefore complexity of the language --- is not needed if one is just using the functionality of an API for data, model and quality engineering to accomplish standard tasks.

\paragraph{Criticism: Static Checking is Hard} Because Python is such a dynamic, general-purpose programming language, it is difficult to statically check the constraints of an API. Since the 2015 release of version 3.5, Python at least supports optional \emph{type hints}, defined in PEP 484 \cite{PEP484}. Still, the authors stress that ``Python will remain a dynamically typed language, and the authors have no desire to ever make type hints mandatory, even by convention''. Thus, taking advantage of type hints requires the use of an external tool such as the static type checker mypy\footnote{http://mypy-lang.org/}. However, of the three ML APIs we observed, only the most recent one, PyTorch, includes type hints. In addition, dependencies, temporal constraints, execution context constraints and ML best practices are beyond the scope of type hints. Even a simple constraint such as "the value must be a float between 0 and 1" cannot be expressed.

The impact of static type checking was quantified (among others) in a study by Fischer and Hanenberg from 2015 \cite{Fischer2015}, where they found the development time of a program in the statically typed JavaScript-superset TypeScript\footnote{\url{https://www.typescriptlang.org}} to be significantly shorter than in the dynamically typed JavaScript itself. We argue that static checking of non-type errors has a similar positive effect.

\subsection{Development Environments}
\label{subsec:ide}
Last but not least, notebooks for Python and other languages, e.g. Jupyter Notebooks\footnote{\url{https://jupyter.org/}}, have become a popular learning and development environment for ML and have recently been integrated into the generic Python IDE PyCharm\footnote{\url{https://www.jetbrains.com/pycharm/}}. 

In a notebook, a program can be split into cells that can be evaluated independently without deleting the results created by other cells. Results, including complex visualizations, are shown directly in the notebook next to the cell. It is possible to include text cells between code cells for documentation purposes and to structure the notebook. This feature also makes notebooks ideal for creating executable teaching materials for novices. Due to their interactivity, they are used in practice as a development environment for ML, especially for data exploration. We claim that the popularity of notebooks, even when used within PyCharm, points to the lack of IDEs that are adapted to the needs of ML development.

\paragraph{Criticism: No Support for Data Engineering} Notebooks and generic Python IDEs such as PyCharm   (henceforth called altogether `\emph{ML tools}') provide no specific support for data engineering. There are no built-in automated analyses of a dataset, as suggested in Sec. \ref{subsec:data_engineering}, or visualizations of them. Developers must rely on external tools or the functionality provided through APIs.

\begin{figure}[h]
    \centering
    \captionsetup{justification=centering}
    \fbox{\includegraphics[width=0.85\linewidth]{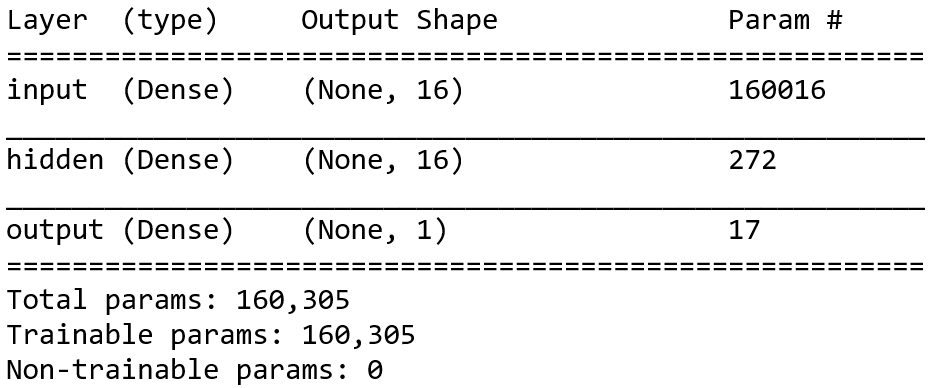}}
    \caption{Output of the summary method (Keras)}
    \label{fig:keras_introspection_summary_output}
\end{figure}

\paragraph{Criticism: No Support for Introspection} While developing a training program, developers need to inspect the current state of data or models. There is no support for this in current ML tools. Instead, developers need to call specific API functions that provide introspection support.
Keras, for example, has a \verb|summary| method that displays the layers of an NN, the shape of their outputs and the number of trainable parameters, which is a good measure of how long training is going to take and whether the model is likely to overfit or underfit. Fig. \ref{fig:keras_introspection_summary_output} shows an example output of this method.

Although introspection is not conceptually part of the learning program, we must mix the two concerns in the notebook environment. A dedicated ML IDE could provide a clearer separation of concerns.

\paragraph{Criticism: No Experiment Management} For experiment management developers must also invent their own solutions or choose external tools. A dedicated IDE could fulfill this need instead and also improve the discoverability of past experiments by automatically comparing the current task and available data to a database, thereby preventing duplicate work.

\section{The Simple-ML Approach}
\label{sec:simpleml}
\begin{figure}[h]
    \centering
    \captionsetup{justification=centering}
    \includegraphics[width=0.8\linewidth]{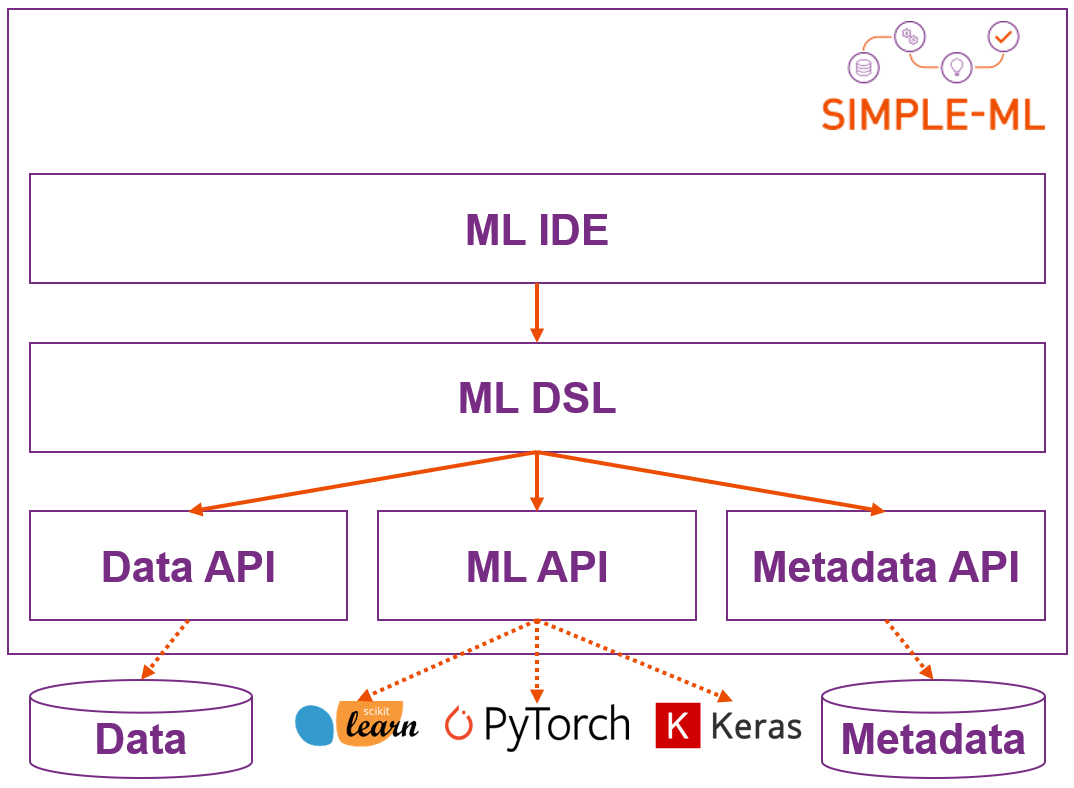}
    \caption{Our Simple-ML Approach}
    \label{fig:simple_ml_approach}
\end{figure}

The Simple-ML project\footnote{https://simple-ml.de/} is dedicated to the development of solutions for the above-mentioned gaps in the state of the art. Its core ideas, summarized in Fig. \ref{fig:simple_ml_approach}, are: (1) the development of a unified ML API, (2) the development of a domain specific language (DSL) for ML, (3) the development of a dedicated ML IDE, and (4) last but not least, as a common basis for (2) and (3), the development of metamodels of the ML domain, ML APIs, and application-specific datasets. 

\paragraph{Towards a Unified ML API}
In its first release, the unified ML API will generalize the functionality of scikit-learn, Keras and PyTorch. Development of an own API lets us (a) ensure the quality of its documentation, and (b) ensure the existence of runtime checks where no static checking is possible. This enables checking of compliance with general, \emph{API-independent} ML best practices once and for all. The unified API will be based on adapters to the different existing APIs and will shield the higher levels from their details. In particular, it will serve as a unified target for code generation by the compiler of the DSL.

\paragraph{A DSL with ML-aware Static Analysis Capabilities} 
The Simple-ML DSL, will provide ML-specific abstractions as first-class language elements, and will make them available via a textual \emph{and} a visual syntax. This will make it easy to learn by developers who want to experiment with ML but are not experienced programmers. In addition, the DSL will statically catch errors related to all the types of constraints identified in Sec. \ref{subsec:api}.


Under the hood the DSL compiler will generate code for a general purpose programming language such as Python, using the implemented adapters to target an individual framework. This avoids locking users into the Simple-ML DSL. If they eventually need the flexibility of a general purpose programming language, they can take the generated code and modify it as desired.

\paragraph{Using Ontologies to Store Metadata} ML-specific error detection will be done based on metadata in the form of ontologies. We will create ontologies that describe ML concepts, such as learning algorithms and quality metrics, to capture information like the hyperparameters of learning algorithms or whether a learning algorithm and a quality metric are compatible. For this we consider improving the MEX ontology \cite{Esteves2015}. 

The ML-specific metadata will be complemented by metadata about available datasets. Each dataset will be associated with a semantic description of its features \cite{Gottschalk2019}. For each feature, we will store metainformation like its semantic category (e.g. the fact that it represents the name of a person), or the associated unit of measurement (e.g. whether a distance is measured in metres or miles). We will also store general statistical properties of a feature, like minimum and maximum values, standard deviation etc.

Using this metadata we can provide the desirable data engineering guidance identified in Sec. \ref{subsec:data_engineering}. For instance, we will be able to quickly detect if data is normalized and normalize it automatically if desired, or provide targeted suggestions for feature combinations based on the semantic categories. For model engineering (Sec. \ref{subsec:model_engineering}) we can also provide guidance, e.g. by suggesting suitable models given the data and the task.
Ontologies can also provide the technical basis for model quality engineering guidance (Sec. \ref{subsec:quality_engineering}). 

\paragraph{An IDE to Tie Everything Together}
Further guidance 
will be provided by an ML-specific IDE that will support the entire ML workflow (Sec. \ref{sec:workflows}) through GUI features. 
The error checking and fixing capabilities of the DSL will be used to display errors and to provide quickfixes while the training program is being written. The ontology-based metadata will be used for elaborate suggestions of preprocessing steps during data engineering (Sec. \ref{subsec:data_engineering}), for suggestions of suitable models during model engineering (Sec. \ref{subsec:model_engineering}), and for experiment management and suggestions of model improvements during  quality engineering (Sec. \ref{subsec:quality_engineering}). 
The GUI of the IDE will also provide introspection (Sec. \ref{subsec:ide}) capabilities, allowing us to eliminate introspection from the user-visible part of the unified ML API, thus making it easier to learn. 

\section{Conclusion}
This paper introduced the concept of guidance as the means offered by an API or tool to facilitate
or enforce correct usage or to help follow a predetermined workflow or established best practices. Guidance speeds up learning, development and debugging. 

We discussed key differences between ML and SE workflows, derived the desirable ML-specific guidance and analyzed how existing APIs, languages and tools used for ML application development fail to provide most of it. As a possible solution, we identified static analysis, meta-modeling, and support by an IDE with an advanced GUI as key SE concepts, whose combined use could provide the missing guidance. Finally, we sketched the Simple-ML project, which will implement this solution in order make ML available as a well-understood tool for software developers.

\begin{acks}
This work was partially funded by the Federal Ministry of Education and Research (BMBF),  Germany under Simple-ML (01IS18054). 
\end{acks}

\bibliographystyle{ACM-Reference-Format}
\bibliography{references}

\end{document}